\documentclass[11pt]{article}
\usepackage{moriond,epsfig,graphics}

\def\Journal#1#2#3#4{{#1} {\bf #2}, #3 (#4)}

\def\NIM{\em Nucl. Instrum. Methods}

\def\PRL{\em Phys. Rev. Lett.}
\def\PRD{{\em Phys. Rev.} D}

\def\be{\begin{equation}}
\def\ee{\end{equation}}
\def\bea{\begin{eqnarray}}
\def\eea{\end{eqnarray}}

\def\dM{\Delta M}

\def\bzb{{\overline{B}{}^0}}
\def\bb{{\overline{B}{}^0}}

\def\bz{{B^0}}

\def\piz{\pi^0}
\def\pip{\pi^+}
\def\pim{\pi^-}

\def\ks{K_S^0}
\def\kl{K_L^0}

\def\jpsi{{J/\psi}}

\def\Dz{$\dz$}
\def\Dt{\Delta t}
\def\Dz{\Delta z}

\def\dE{\Delta E}

\def\mb{M_{\rm bc}}

\newcommand{\Btag}{B_{\rm tag}}
\newcommand{\zcp}{z_{CP}}
\newcommand{\ztag}{z_{\rm tag}}

\newcommand{\sinbb}{{\sin2\phi_1}}

\def\dM{{\Delta m_d}}

\begin{document}
\vspace*{4cm}
\title{IMPROVED MEASUREMENT OF \boldmath $CP$ ASYMMETRY\\IN THE NEUTRAL $B$ MESON SYSTEM}

\author{T. Higuchi
 \footnote{The author is a Research Fellow of the Japan Society for the Promotion of Science.}
\\(for the Belle collaboration)}

\address{Department of Physics, Faculty of Science, the University of Tokyo,\\
7-3-1 Hongo, Bunkyo-ku, Tokyo 113-0033, Japan}

\maketitle\abstracts{
We present an improved measurement of the standard model $CP$ violation
parameter $\sinbb$ (also known as $\sin2\beta$) based on a sample of $45\times10^6$
$B\overline{B}$ pairs collected at the $\Upsilon(4S)$ resonance
with the Belle detector at the KEKB asymmetric-energy $e^+e^-$ collider.
One neutral $B$ meson is reconstructed in the $\jpsi\ks$, $\psi(2S)\ks$,
$\chi_{c1}\ks$, $\eta_c\ks$, $\jpsi K^{*0}$, or $\jpsi\kl$ $CP$-eigenstate
decay channel and the flavor of accompanying $B$ meson is identified from its
decay products.  From the asymmetry in the distribution of the time intervals
between the two $B$ meson decay points, we obtain
$\sinbb=0.82\pm0.12\mbox{(stat)}\pm0.05\mbox{(syst)}$.
The result is preliminary.
}

In the Standard Model (SM), $CP$ violation arises from an
irreducible complex phase in the weak interaction quark mixing matrix
(CKM matrix)\cite{bib:ckm}.
In particular, the SM predicts a $CP$ violating asymmetry
in the time-dependent rates for $\bz$ and $\bb$ decays to a common
$CP$ eigenstate, $f_{CP}$,
with negligible corrections from strong interactions\cite{bib:sanda}:
\be
A(t) \equiv \frac{\Gamma(\bb\to f_{CP}) - \Gamma(\bz\to f_{CP})}
{\Gamma(\bb\to f_{CP}) + \Gamma(\bz\to f_{CP})} = -\xi_f \sinbb \sin\dM t,
\ee
where $\Gamma(\bz,\bb \to f_{CP})$ is the decay rate for a $\bz$ or $\bb$
to $f_{CP}$ dominated by $b\to c\overline{c}s$ transition
at a proper time $t$ after production, $\xi_f$ is the $CP$ eigenvalue of
$f_{CP}$, $\dM$ is the mass difference between the two $\bz$ mass eigenstates,
and $\phi_1$ is one of the three interior angles of the CKM unitarity triangle,
defined as $\phi_1 \equiv \pi-\arg(-V_{tb}^*V_{td}/-V_{cb}^*V_{cd})$.

In this paper, we report an improved measurement of $\sinbb$ using
$45\times 10^6$ $B\overline{B}$ pairs (42~fb$^{-1}$) collected with the Belle detector\cite{bib:belle} at the $\Upsilon(4S)$ resonance in collisions of 8.0~GeV $e^-$ to
3.5~GeV $e^+$ at KEKB\cite{bib:kekb}, where two $B$ mesons reside in a coherent
$p$-wave state until one of
them decays.  The decay of one of the $B$ mesons to a self-tagging state,
$f_{\rm tag}$, {\it i.e.}, a final state that distinguishes $\bz$ and
$\bb$, at a time $t_{\rm tag}$ projects the accompanying meson onto the
opposite $b$-flavor at that time; this meson decays to $f_{CP}$ at time
$t_{CP}$.  The $CP$ violation manifests itself as an asymmetry $A(\Dt)$,
where $\Dt$ is the proper time interval $\Dt \equiv t_{CP}-t_{\rm tag}$.
At KEKB, the $\Upsilon(4S)$ resonance is produced with a boost of $\beta\gamma=0.425$
along the electron beam direction ($z$ direction).  Because
the $\bz$ and $\bb$ mesons are nearly at rest in the $\Upsilon(4S)$ center
of mass system (cms), $\Dt$ can be determined as $\Dt \simeq \Dz/(\beta\gamma)c$,
where $\Dz$ is the $z$ distance
between the $f_{CP}$ and $f_{\rm tag}$ decay vertices, $\Dz \equiv \zcp-\ztag$.
The average value for $\Dz$ is approximately 200 $\mu$m.

\begin{figure}[t]
\hspace{0.1cm}
\begin{minipage}{7.8cm}
\begin{center}
\hspace{0.008cm}
\psfig{figure=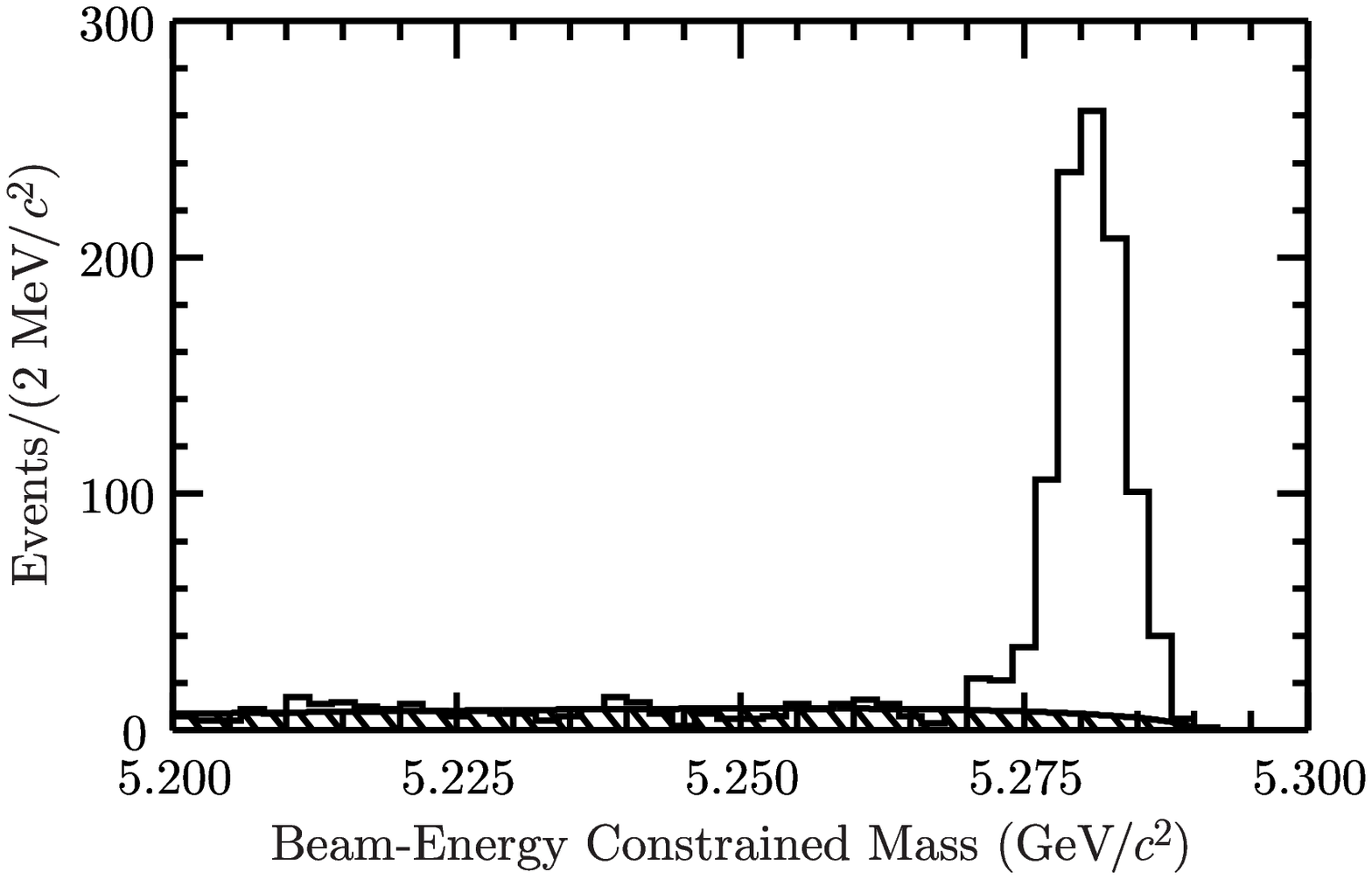,width=6.9cm,height=2.8cm}
\\
\psfig{figure=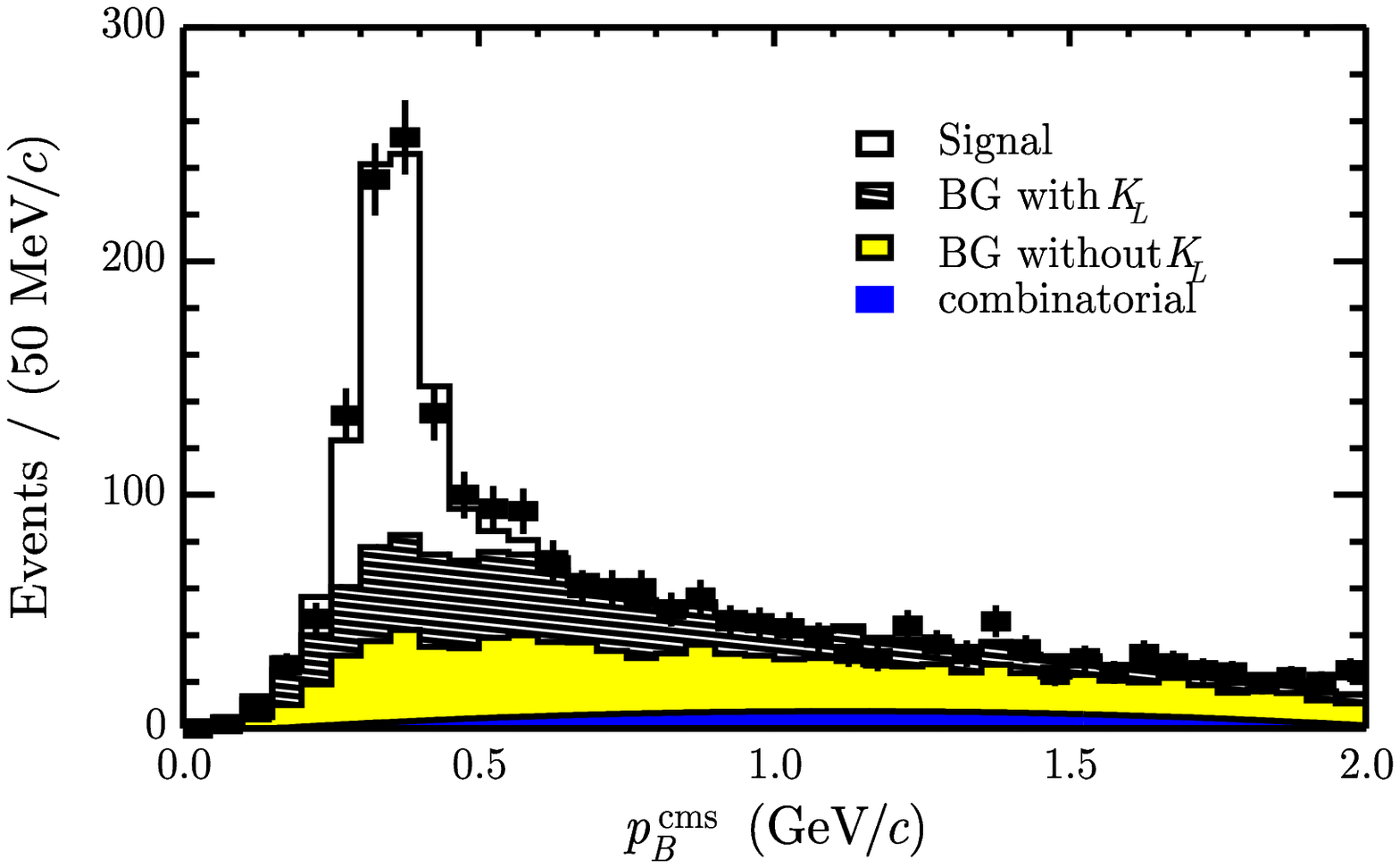,width=6.7cm,height=2.7cm}
\\
\vspace*{-0.1cm}
\caption{(Top) the beam-energy constrained mass distribution for all decay modes
combined other than $\jpsi\kl$.
(Bottom) the $p_B^{\rm cms}$ distribution for $\bz\to\jpsi\kl$ candidates with the results of the fit.
}
\label{fig:mbc}
\end{center}
\end{minipage}
\hspace*{0.1cm}
\begin{minipage}{7.6cm}
\hspace{0.3cm}
\begin{center}
\begin{minipage}{7.6cm}
\footnotesize
Table 1: The numbers of observed candidates ($N_{\rm ev}$) and the estimated
background ($N_{\rm bkg}$) in the signal region for each $f_{CP}$ mode.
\vspace*{0.3cm}
\end{minipage}
\label{tab:number}
\vspace{0.4cm}
\begin{tabular}{|l|r|r|}
\hline
Mode & $N_{\rm ev}$ & $N_{\rm bkg}$ \\
\hline
$\jpsi(\ell^+\ell^-)\ks(\pi^+\pi^-)$       & 636 & 31.2 \\
$\jpsi(\ell^+\ell^-)\ks(\pi^0\pi^0)$       & 102 & 20.8 \\
$\psi(2S)(\ell^+\ell^-)\ks(\pi^+\pi^-)$    &  49 &  2.4 \\
$\psi(2S)(\jpsi\pi^+\pi^-)\ks(\pi^+\pi^-)$ &  57 &  4.3 \\
$\chi_{c1}(\jpsi\gamma)\ks(\pi^+\pi^-)$    &  34 &  2.3 \\
$\eta_c(K^+K^-\pi^0)\ks(\pi^+\pi^-)$       &  39 & 11.1 \\
$\eta_c(\ks K^-\pi^+)\ks(\pi^+\pi^-)$      &  33 &  8.9 \\
\hline
$\jpsi(\ell^+\ell^-) K^{*0}(\ks\pi^0)$     &  55 &  6.0 \\
\hline
$\jpsi(\ell^+\ell^-) \kl$                  & 767 & 307 \\
\hline
\end{tabular}
\end{center}
\end{minipage}
\vspace*{-.5cm}
\end{figure}

We reconstruct $\bz$ decays to the following $CP$ eigenstates
\footnote{Throughout this paper, when a decay mode is quoted,
the inclusion of the charge conjugation mode is implied.}:
$\jpsi\ks$, $\psi(2S)\ks$, $\chi_{c1}\ks$, $\eta_c\ks$ for $\xi_f = -1$ and
$\jpsi\kl$ for $\xi_f$ = +1.  We also use $\bz\to\jpsi K^{*0}$ decays where
$K^{*0} \to \ks\piz$.  Here the final state is a mixture of even and odd
$CP$, depending on the relative orbital angular momentum of the $\jpsi$
and $K^{*0}$.  We find that the final state is primarily $\xi_f$; the
$\xi_f=-1$ fraction is $0.19\pm0.04\mbox{(stat)}\pm0.04\mbox{(syst)}$\cite{bib:itoh}.
For reconstructed $B$ candidates except $\jpsi\kl$,
we identify $B$ decays using the
energy difference $\dE\equiv E_B^{\rm cms}-E_{\rm beam}^{\rm cms}$ and the
beam-energy constrained mass
$\mb \equiv \sqrt{(E_{\rm beam}^{\rm cms})^2-(p_B^{\rm cms})^2}$, where
$E_{\rm beam}^{\rm cms}$, $E_B^{\rm cms}$, and $p_B^{\rm cms}$ are the beam
energy, the energy, and the momentum of the reconstructed $B$ candidate in the
cms, respectively.
Figure \ref{fig:mbc} (top) shows the $\mb$ distributions for all
$\bz$ candidates except for $\bz\to\jpsi\kl$ that are found in the $\dE$ signal
region.
Table 1 lists the
numbers of observed candidates ($N_{\rm ev}$) and the background ($N_{\rm bkg}$)
estimated by extrapolating the rate in the non-signal $\dE$ vs $\mb$ region
into the signal region.
Candidate $\bz\to\jpsi\kl$ decays are selected by requiring
electromagnetic calorimeter
(ECL) and/or $\kl$ and muon detector
(KLM) hit patterns that are consistent with the presence of a shower
induced by a neutral hadron.
The centroid of the shower is required to be
in a $45^\circ$ cone centered on the $\kl$ direction that is inferred from
two-body decay kinematics and the measured four-momentum of the $\jpsi$.
Figure \ref{fig:mbc} (bottom) shows the $p_B^{\rm cms}$ distribution,
calculated with the $\bz\to\jpsi\kl$ two-body decay hypothesis.  The histograms
are the results of a fit to the signal and background distributions.
There are 767 entries in total in the region $0.20\le p_B^{\rm cms}\le0.40$~GeV/$c$ with KLM clusters and in $0.20\le p_B^{\rm cms}\le0.40$~GeV/$c$ with
clusters in the ECL only.
The fit finds a signal purity of 60\%.
The reconstruction and selection criteria for used channels
are described elsewhere\cite{bib:cpv}.

Leptons, charged pions, kaons, and $\Lambda$ baryons that are not associated
with a reconstructed $CP$ eigenstate decay are used to identify the $b$-flavor
of the accompanying $B$ meson;
high momentum leptons from $b\to c\ell^-\overline{\nu}$,
lower momentum leptons from $c\to s\ell^-\overline{\nu}$,
charged kaons and $\Lambda$ baryons from $b\to c \to s$,
fast pions from $\bz\to D^{(*)-}$($\pi^+,\rho^+,a_1^+$, etc.), and
slow pions from $D^{*-}\to \overline{D}{}^0\pi^-$.
Using those tracks, two parameters, $q$ and $r$, are assigned to an event.
The first, $q$, has the discrete values $q=\pm1$ that is $+1(-1)$
when $\Btag$ is likely to be a $\bz$($\bb$), and the parameter $r$ is an
event-by-event flavor-tagging dilution factor ranging from $r=0$ for no flavor
discrimination to $r=1$ for unambiguous flavor assignment.  It is used only
to sort data into six intervals of $r$, according to flavor purity;
the wrong-tag probabilities, $w_l~(l=1,6)$, for the final fit are determined
directly from data.
The decay to exclusively reconstructed self-tagged channels
are utilized to obtain $w_l$ using time-dependent $\bz$-$\bb$ mixing
oscillation: $(N_{\rm OF}-N_{\rm SF})/(N_{\rm OF}+N_{\rm SF}) = (1-2w_l)\cos(\dM\Dt)$.  Here $N_{\rm OF}$ and $N_{\rm SF}$ are the numbers of opposite
and same flavor events.
The total effective tagging efficiency is $\sum_{l=1}^6f_l(1-2w_l)^2 = 0.270\pm0.008\mbox{(stat)}^{+0.006}_{-0.009}\mbox{(syst)}$,
where $f_l$ is the event fraction for each $r$ interval.

The vertex position for $f_{CP}$ is reconstructed using leptons from $\jpsi$
decay and that for $f_{\rm tag}$ is obtained with well reconstructed tracks
not assigned to $f_{CP}$.  Tracks that form a $\ks$ invariant mass are not used.
Each vertex position is required to be consistent with the interaction
point profile smeared in the $r$-$\phi$ plane by the $B$ meson decay length.
The requirement enables us to determine a vertex even with a single track.
The fraction of such vertices is about 10\% for $\zcp$ and 30\% for $\ztag$.
A proper-time interval resolution, $R_{\rm sig}(\Dt)$,
consists of a convolution of four components:
detector resolution for $\zcp$ and $\ztag$, shift in the $\ztag$ reconstruction
due to secondary tracks originated from charmed hadrons such as $D^+$ and $D^0$, and
smearing effect due to kinematic approximation in converting
$\Dz$ to $\Dt$.
We find broad outlier components in $\Dz$ distributions due to
mis-reconstruction, which are represented by Gaussian distributions.
We simultaneously
determine ten resolution parameters from data with a fit of neutral
and charged $B$ meson lifetimes\cite{bib:lifetime}
and obtain a $\Dt$ resolution of $\sim1.56$~ps (rms).
The width and the fraction of the outlier component are determined to be
$36^{+5}_{-4}$~ps, and $(0.06^{+0.03}_{-0.02})\%$ or $(3.1\pm0.4)\%$
(multiple- or single-track case).

\begin{figure}[t]
\hspace{0.1cm}
\begin{minipage}{7.4cm}
\begin{center}
\psfig{figure=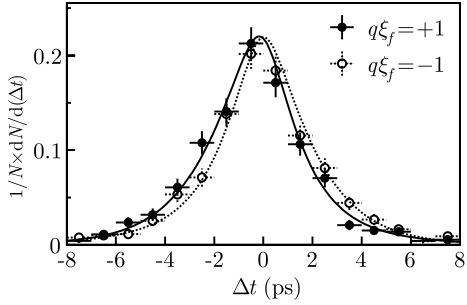,height=4.0cm}
\vspace*{-0.2cm}
\caption{$\Dt$ distributions for the events with $q\xi_f=+1$ (solid points)
and $q\xi_f=-1$ (open points).  The results of the global fit with
$\sinbb=0.82$ are shown as solid and dashed curves, respectively.}
\label{fig:cpfit}
\end{center}
\end{minipage}
\hspace*{0.1cm}
\begin{minipage}{8.0cm}
\hspace{0.3cm}
\begin{center}
\begin{minipage}{8.0cm}
\footnotesize
Table 2: The values of $\sinbb$ for various subsamples (statistical errors only).
\vspace*{0.3cm}
\end{minipage}
\label{tab:check}
\begin{tabular}{|l|c|}
\hline
Sample & $\sinbb$ \\
\hline
$f_{\rm tag}=\bz$ $(q=+1)$                        & $0.60\pm0.19$ \\
$f_{\rm tag}=\bb$ $(q=-1)$                        & $0.99\pm0.16$ \\
$\jpsi\ks(\pip\pim)$                              & $0.67\pm0.18$ \\
$(c\overline{c})\ks$ except $\jpsi\ks(\pip\pim)$  & $0.88\pm0.31$ \\
$\jpsi\kl$                                        & $1.14\pm0.23$ \\
$\jpsi K^{*0}(\ks\pi^0)$                          & $1.62\pm1.10$ \\
\hline
All                                               & $0.82\pm0.12$ \\
\hline
\end{tabular}
\end{center}
\end{minipage}
\end{figure}

After vertexing we find 766 events with $q=+1$ flavor tags and 784 events
with $q=-1$.  Figure \ref{fig:cpfit} shows the observed $\Dt$ distributions
for the $q\xi_f=+1$ (solid points) and $q\xi_f=-1$ (open points) event samples.
The asymmetry between two distributions demonstrates the violation of the
$CP$ symmetry.
We determine $\sinbb$ from an unbinned maximum-likelihood fit to the observed
$\Dt$ distributions.  The probability density function (pdf) expected
for the signal distribution is given by
\be
{\cal P}_{\rm sig}(\Dt,q,w_l,\xi_f) =
\frac{e^{-|\Dt|/\tau_\bz}}{2\tau_\bz}[1-q\xi_f(1-2w_l)\sinbb\sin\dM\Dt],
\ee
where we fix the $\bz$ lifetime and mass difference at their world average
values\cite{bib:pdg}.
Each pdf is convolved with the appropriate $R(\Dt)$
to determine the likelihood value for each event as a function of $\sinbb$:
\begin{eqnarray}
P_i &=& (1-f_{\rm ol})\int \Bigl[ f_{\rm sig}{\cal P}_{\rm sig}(\Dt',q,w_l,\xi_f)R_{\rm sig}(\Dt-\Dt') \nonumber \\
&& +\; (1-f_{\rm sig}){\cal P}_{\rm bkg}(\Dt')R_{\rm bkg}(\Dt-\Dt')\Bigr] d\Dt'
+ f_{\rm ol}P_{\rm ol}(\Dt),
\end{eqnarray}
where $f_{\rm sig}$ is the signal probability calculated
as a function of $p_B^{\rm cms}$ for $\jpsi\kl$ and of $\dE$ and $\mb$ for
other modes.
${\cal P}_{\rm bkg}(\Dt)$ is a pdf for combinatorial background events
that is modeled as a sum of exponential and prompt components.  It is convolved
with a sum of two Gaussians, $R_{\rm bkg}$.
$P_{\rm ol}$ and $f_{\rm ol}$ are the pdf and the fraction for the outlier
component.
The only free parameter in the final fit is $\sinbb$, which is determined by
maximizing the likelihood function $L=\prod_iP_i$, where the product is over
all events.  The preliminary result of the fit is
\[
\sinbb = 0.82\pm 0.12\mbox{(stat)}\pm0.05\mbox{(syst)} .
\]

The systematic error is dominated by uncertainties due to effects of the tails
of the vertex distribution ($\pm$0.030).  Other significant
contributions come from uncertainties
(a) in $w_l$ ($^{+0.024}_{-0.026}$);
(b) in the resolution function parameters ($^{+0.022}_{-0.019}$);
(c) in the $\jpsi\kl$ background fraction ($^{+0.014}_{-0.015}$).
The errors introduced by uncertainties in $\dM$ and $\tau_\bz$ are 0.01 or less.

A number of checks on the measurement are performed.  Table 2
lists the results obtained by applying the same analysis to various subsamples.
All values are statistically consistent with each other.
The result is unchanged
if we use the $w_l$'s determined separately for $f_{\rm tag}=\bz$ and $\bzb$.
A fit to the non-$CP$ eigenstate self-tagged modes $\bz\to D^{(*)-}\pi^+$,
$D^{*-}\rho^+$, and $\jpsi K^{*0}(K^+\pi^-)$, where no asymmetry is expected,
yields $0.05\pm0.04$.
\\

We also measure $CP$ violating asymmetry in $\bz\to\eta'\ks$ decays
based on $45\times10^6$ $B\overline{B}$ pairs.
Numbers of fully reconstructed events
for $\eta(\gamma\gamma)\pi\pi\ks$ and for $\rho\gamma\ks$
are $27.7^{+6.2}_{-5.5}$ and $45.4^{+8.6}_{-7.9}$, respectively.
Flavor of the accompanying $B$ meson is identified from its decay products.
The decay rate has a time dependence given by
\be
{\cal P}_{\rm sig}(\Dt,q,w_l) = \frac{e^{-|\Dt|/\tau_\bz}}{4\tau_\bz}\Bigl\{1+q(1-2w_l)[S\sin(\dM\Dt)+C\cos(\dM\Dt)]\Bigr\}.
\ee
From the asymmetry in the $\Dt$ distribution,
we obtain $S=0.27^{+0.54}_{-0.55}\mbox{(stat)}\pm0.07\mbox{(syst)}$ and
$C=0.12\pm0.32\mbox{(stat)}\pm0.07\mbox{(syst)}$.

\end{document}